\documentclass[apjl,final,article]{emulateapj}
 
\bibpunct{(}{)}{;}{a}{}{,} 
\newcommand{\bm}{\boldmath}
\newcommand{\bvc}[1]{\mbox{\bm $#1$}}
\newcommand{\caii}{Ca\,{\small II}}

\begin{document}


\title{Evidence of Shock-Driven Turbulence in the Solar Chromosphere}

\author{K.P. Reardon\altaffilmark{1,2}}
\author{F. Lepreti\altaffilmark{3}}
\author{V. Carbone\altaffilmark{3,4}}
\author{A. Vecchio\altaffilmark{1,3}}
\altaffiltext{1}{INAF -- Osservatorio Astrofisico di Arcetri,
              Largo E. Fermi, 5,
              50125 Firenze, Italy,
              kreardon@arcetri.astro.it}
\altaffiltext{2}{Visiting Astronomer,
              National Solar Observatory,
              Sunspot, NM, 88349}
\altaffiltext{3}{Dipartimento di Fisica, Universit\`a della  Calabria,
              Via P. Bucci 31/C,
              87036 Rende (CS), Italy}
\altaffiltext{4}{LICRYL, INFM/CNR, Via P. Bucci 31/C, I-87036 Rende (CS), Italy}

\submitted{Received: 2008 March 18; accepted 2008 July 11; 2008 August 12}
\journalinfo{The Astrophysical Journal, 683:L207--L210, 2008 August 20}

\begin{abstract}
We study the acoustic properties of the solar chromosphere in the high-frequency regime using a time sequence of velocity measurements in the chromospheric \caii~854.2 nm line taken with the Interferometric Bidimensional Spectrometer (IBIS). We concentrate on quiet-Sun behavior, apply Fourier analysis, and characterize the observations in terms of the probability density functions (PDFs) of velocity increments. We confirm the presence of significant oscillatory fluctuation power above the cutoff frequency and find that it obeys a power-law distribution with frequency up to our 25~mHz Nyquist limit. The chromospheric PDFs are non-Gaussian and asymmetric and they differ among network, fibril, and internetwork regions. This suggests that the chromospheric high-frequency power is not simply the result of short-period waves propagating upward from the photosphere but rather is the signature of turbulence generated within the chromosphere from shock oscillations near the cutoff frequency. The presence of this pervasive and broad spectrum of motions in the chromosphere is likely to have implications for the excitation of coronal loop oscillations.
\end{abstract}
\keywords{Sun: chromosphere, turbulence,   shock waves, Sun: photosphere}

\section{Introduction}
\label{sec:intro}

The role played by high-frequency acoustic oscillations in the heating of the solar chromosphere and corona has been addressed in numerous studies. Because acoustic oscillations above the cutoff frequency freely propagate upward, they have long been considered candidates for transporting to the outer solar atmosphere some of the abundant mechanical energy generated below the photosphere \citep[see review by][]{1996SSRv...75..453N}. 

The primary test for this theory has been to measure the high-frequency oscillations in the photosphere and to estimate whether they contain sufficient acoustic flux to balance the chromospheric and coronal losses. 
For example, \cite{2002A&A...395L..51W} and \cite{ 2007SoPh..242....9A} have looked for power in velocities and intensities measured in photospheric lines. Other authors  \citep{2005A&A...430.1119D, 2004ApJ...617L..89D} have used {\em Transition Region and Coronal Explorer} ({\em TRACE}) UV continuum images to measure power spectra of the intensity fluctuations in the upper photosphere. \citet{2006ApJ...646..579F} also used {\em TRACE} but deduced the upward acoustic flux through comparison with hydrodynamic simulations, finding that the energy carried by high-frequency waves is insufficient to balance the chromospheric losses. This analysis was recently repeated by \citet{2007PASJ...59S.663C}
using the Solar Optical Telescope aboard {\em Hinode}.

Chromospheric oscillations have generally been measured using the \caii~H\,and\,K and infrared triplet lines. Imaging measurements using broad filters in H\,and\,K are mostly dominated by the photospheric contribution within the filter bandpass and do not provide a clear indication of chromospheric behavior.
Traditional slit spectroscopy gave evidence of significant oscillatory power above the acoustic cutoff frequency \citep[e.g.][]{1963AnAp...26..368E,1974ApJ...189..359L, 1978A&A....70..345C, 1981A&A....97..310M, 1990A&A...228..506D}. The breakthrough simulation by \citet{1997ApJ...481..500C} of the \caii\ H spectrogram sequence of 
\citet{1993ApJ...414..345L} 
demonstrated that photospheric oscillations produce chromospheric shocks through upward propagation, but the role they play in chromospheric heating and their energetics are still under debate \citep[e.g.][]{2007ApJ...671.2154K}.

In this Letter we analyze a time series of chromospheric and photospheric velocity images, obtained
at high spatial resolution with the Interferometric Bidimensional Spectrometer 
\citep[IBIS;~][]{2006SoPh..236..415C, paper2}. We investigate the scaling behavior of high-frequency velocity fluctuations in chromospheric regions with different magnetic topologies, to study the basic properties of chromospheric turbulence. In Sec.~\ref{sec:obs} we describe the observations and discuss the velocity power spectra, in Sect.~\ref{sec:turbulence} we present the analysis of statistical properties of velocity increments, 
and in in Sec~\ref{sec:conclusions} our conclusions are summarized.

\section{Observations}
\label{sec:obs}

\begin{figure}[htbp]
\begin{center}
\epsscale{0.75}
\plotone{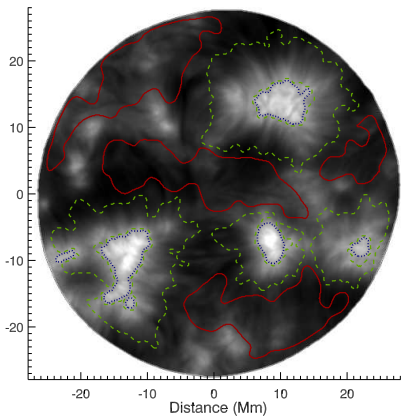}
\caption{Line-center intensity in the \caii~854.2 averaged over the  duration of the observations. The contours outline the masks for the network ({\it blue dotted contours}), fibril ({\it green dashed contours}), and internetwork ({\it red solid contours}) regions. The substantial unclassified regions, primarily between fibril and internetwork regions, are not analyzed.}
\label{fig:fov}
\end{center}
\end{figure}

We analyze a dataset obtained with IBIS at the Dunn Solar Telescope of the National Solar Observatory on 2004 June 2 covering an 80\arcsec~diameter field in the quiet Sun at disk center. These data consist of imaging spectral scans made through the Fe\,{\small I}~709.0~nm (height of formation $\sim$200 km above $\tau_{500}=1$) and \caii~854.2~nm (height of formation $\sim$1000~km) lines with a spectral sampling of 32 and 80~m\AA, respectively. The overall cadence for the complete scan was 19 seconds and the scan was repeated 175 times for a total period of 55 minutes. To eliminate a significant source of noise in the assembled spectra, the narrowband images were carefully destretched using broadband images obtained simultaneously with a common shutter.
More detail is given by \citet{2006A&A...450..365J} and \citet{2007A&A...461L...1V}. 

Doppler velocities were measured by fitting a second-order polynomial to the spectral positions around the core of the line after Fourier interpolation onto a finer resolution grid and by determining the wavelength shift of its minimum. We have performed tests of the accuracy of this approach by adding random noise, consistent with the photon statistics of the data, onto an averaged reference spectrum and found that, for this spectral sampling, the precision of the line core position is better than 1.4 m$\AA$ or 60 m~s$^{-1}$ for the Fe\,{\small I}~709.0 nm line and better than 10 m$\AA$ or 350 m~s$^{-1}$ for the \caii~854.2 line.

We separate several chromospheric regions in the field of view on the basis of differences within the maps of \caii\ 854.2 line core intensity and the maps of Fourier power of the \caii\ line core velocity in the 3-minute ($4.2\!-\!6.6$~mHz) and 5-minute regimes ($3.0\!-\!3.6$~mHz) following \citet{2007A&A...461L...1V}. 
We add the power constraint because \citet{2008A&A...480..515C} showed that the traditional division into network, internetwork, and intermediate regions based on apparent brightness alone \citep[e.g.][]{2001A&A...379.1052K} does not effectively separate areas with distinctly different chromospheric dynamics, in particular the presence or absence of fibrils in the intermediate brightness class.
The three areas we thus define --- namely, network (high-intensity, high 5-minute power), fibril (low-intensity, low 3-minute power), and internetwork (low-intensity, high 3-minute power) --- are shown in Fig. \ref{fig:fov}.

\begin{figure}[htbp]
\begin{center}
\epsscale{0.9}
\plotone{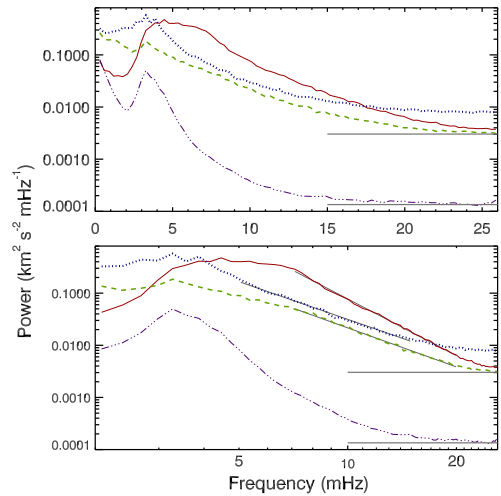}
\caption{Power spectra of the chromospheric velocities for the different types of areas outlined in Fig.~\ref{fig:fov}, and
plotted on both log-linear ({\it top}) and log-log ({\it bottom}) scales.
The power spectrum for the photospheric Fe\,{\small I}~7090 nm line averaged over all three regions ({\it purple dot-dashed}) is also shown.
The horizontal gray lines show the noise level for the chromospheric and photospheric power spectra as estimated in Sec.~\ref{sec:obs}. 
The linear fits to the high-frequency tails are shown in gray on the log-log plots.}
\label{fig:powsp}
\end{center}
\end{figure}

For each pixel within these three selected areas we performed a one-dimensional Fourier transform of the Doppler velocity of \caii~854.2 nm line and averaged these over that area. For the photospheric Fe\,{\small I}~709.0 nm line the differences in the power spectra among these three masks are not significant, and we calculate a single power spectrum for all three areas. Figure~\ref{fig:powsp} shows the resulting power spectra, which display characteristic behavior with a sharp peak around 3.3~mHz for the Fe\,{\small I}~709.0 line, whereas the \caii~854.2 nm line shows a broad plateau over $3\!-\!5$ mHz in the internetwork, demonstrating the well-known predominance of 3-minute oscillations there.

In this Letter we emphasize the presence of high-frequency tails in the average power spectra of each type of region. They lie well above our noise estimations from the photon statistics (indicated by the horizontal lines in Fig.~\ref{fig:powsp}), nearly out to the Nyquist frequency of 26~mHz.
Such high-frequency tails were already evident in traditional spectrographic observations by, e.g.,\ \citet[Figs.~4 and 5]{1963AnAp...26..368E}, \citet[Fig.~4]{1966ApJ...143..917O}, \citet[Fig.~1]{1981SoPh...69..233W}, \citet[Fig.~2]{1990A&A...228..506D}, but our data show them particularly well due to a high signal-to-noise and good statistics from the two-dimensional spectroscopy. 

The bottom panel of Fig.~\ref{fig:powsp} plots the same power spectra on log-log scales. The nearly linear trend of the high-frequency tails suggests power-law behavior for all three chromospheric areas. We fit a line to these tails over the range $7\!-\!20$~mHz ($5\!-\!15$~mHz for the network, where the power spectra peaks at lower frequencies) and find slopes of $-2.4$,$-2.5$, and $-3.6$ for the network, fibril, and internetwork regions respectively. A chi-squared goodness-of-fit test using these linear fits shows that the tails of the power spectra are well represented as a linear trend. A $1/f$ noise source would also display a power law distribution, but with a slope of unity. 

The decreasing density with height results in a proportional increase in the velocity amplitude and would also produce an enhancement of the high-frequency power in the chromosphere compared to the photosphere. However, a process based on a simple density scaling would be expected to operate equally at all frequencies and for all chromospheric regions (except for a variation in magnitude due to differing density gradients).  Instead, the form of the power spectra changes between the photosphere and the chromosphere, with the appearance of the power law trend at high frequencies. Furthermore, the slope of this high-frequency tail varies significantly in different chromospheric regions, which implies that the chromospheric structures play a role in modifying the observed velocities beyond a simple scaling of the underlying photospheric motions.
We also note that the comparisons between power spectra from lines arising at different heights made by \citet{1967IAUS...28..293N} and \citet{1981SoPh...69..233W} clearly show that the power spectra undergo an abrupt change near the base of the chromosphere.

\section{Turbulence Characterization: Velocity Increments}
\label{sec:turbulence}

  \begin{figure*}
   \centering
   \includegraphics[width=5.11cm,trim = 15mm 7mm 12mm 13mm, clip]{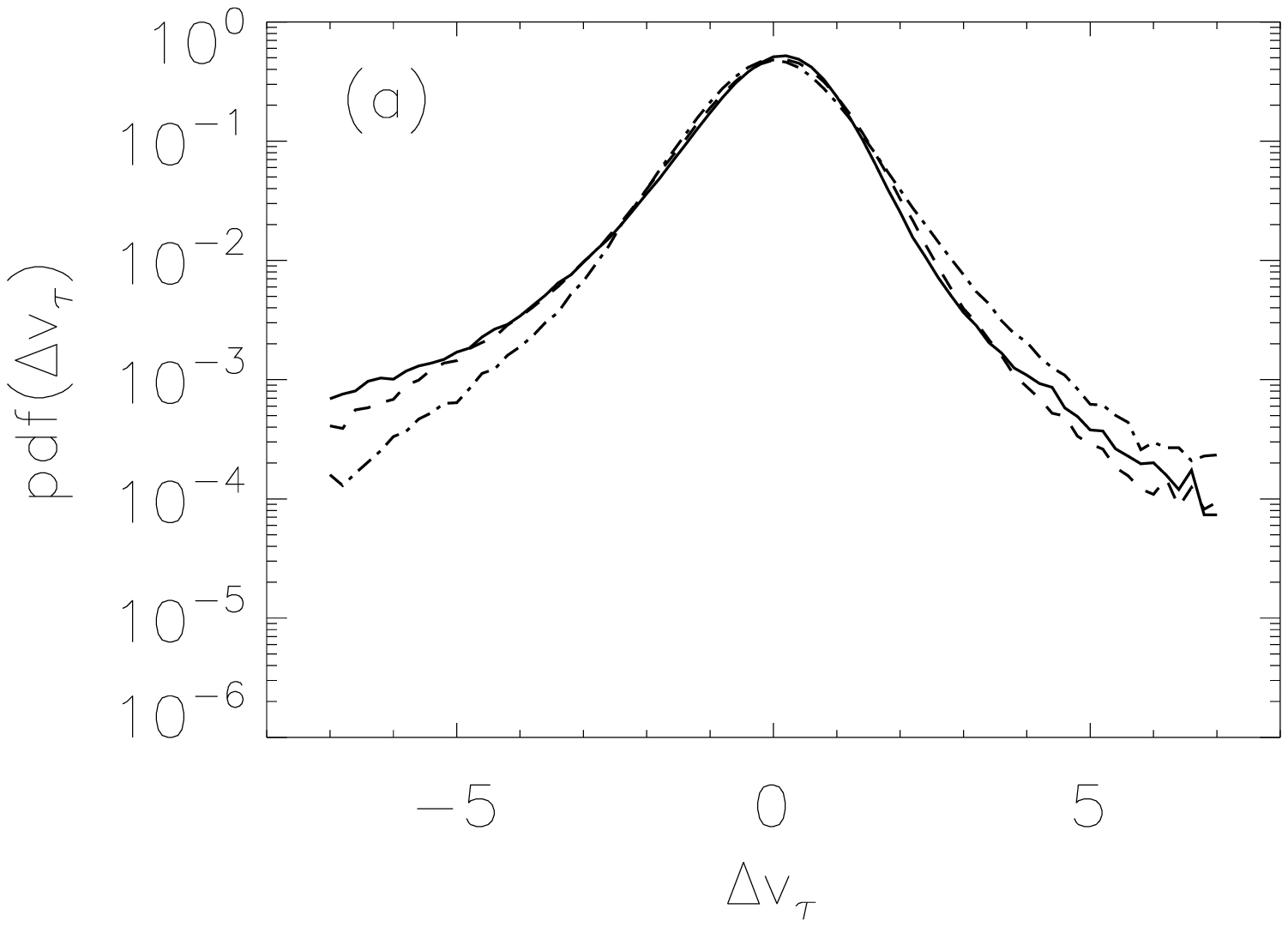}
   \includegraphics[width=4.2cm,trim = 42mm 7mm 12mm 13mm, clip]{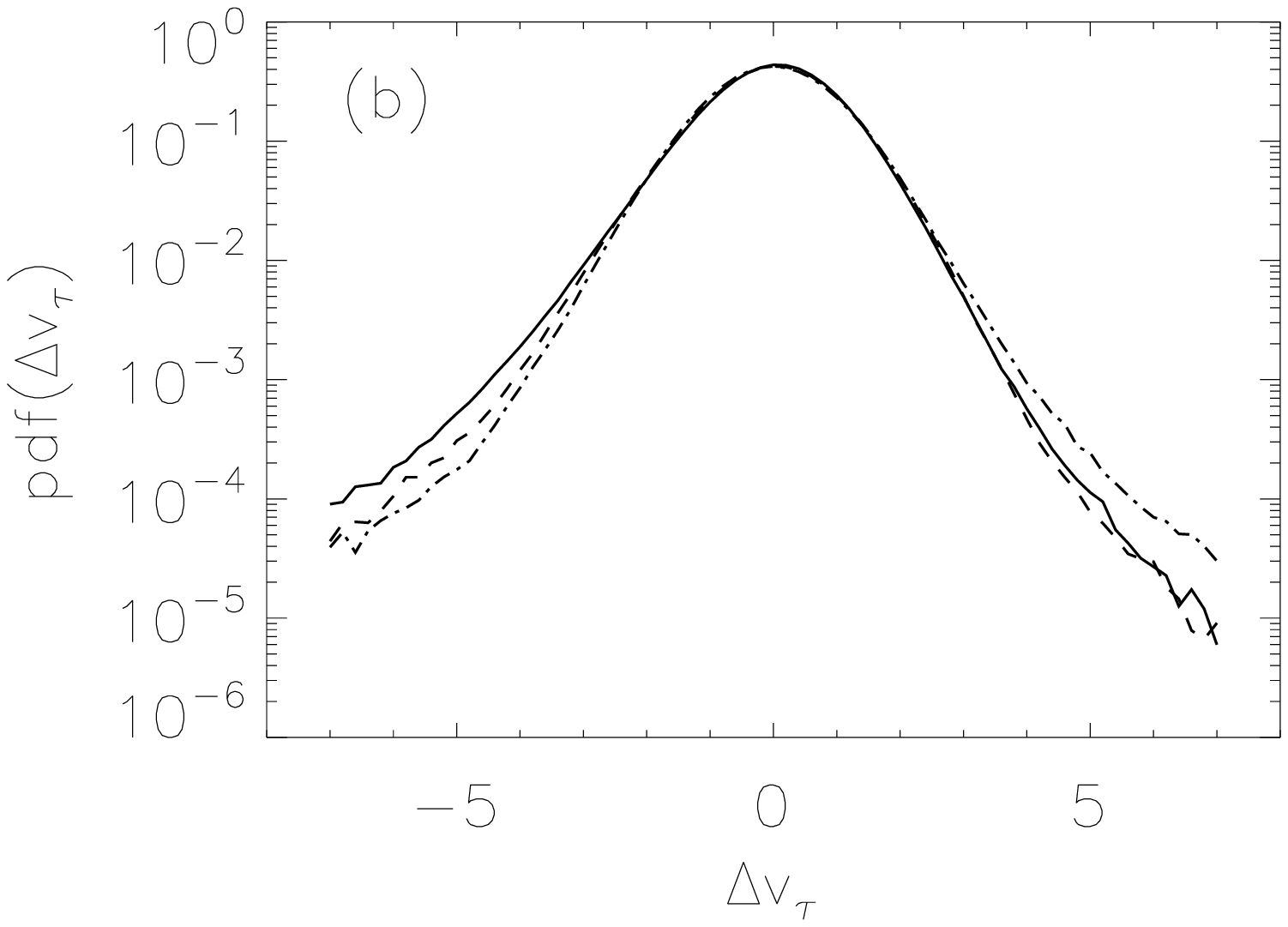}
   \includegraphics[width=4.2cm,trim = 42mm 7mm 12mm 13mm, clip]{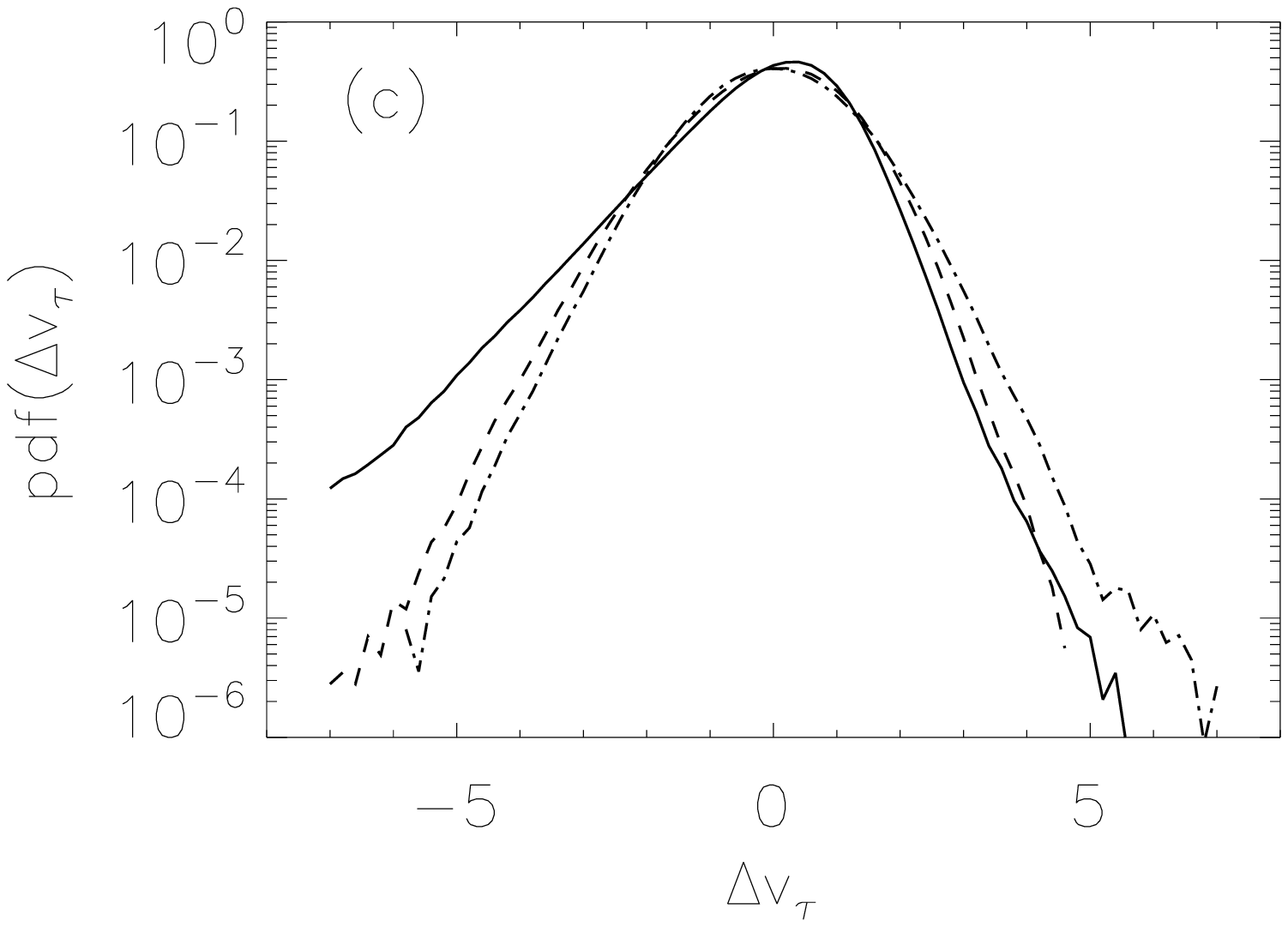}
   \includegraphics[width=4.2cm,trim = 42mm 7mm 12mm 13mm, clip]{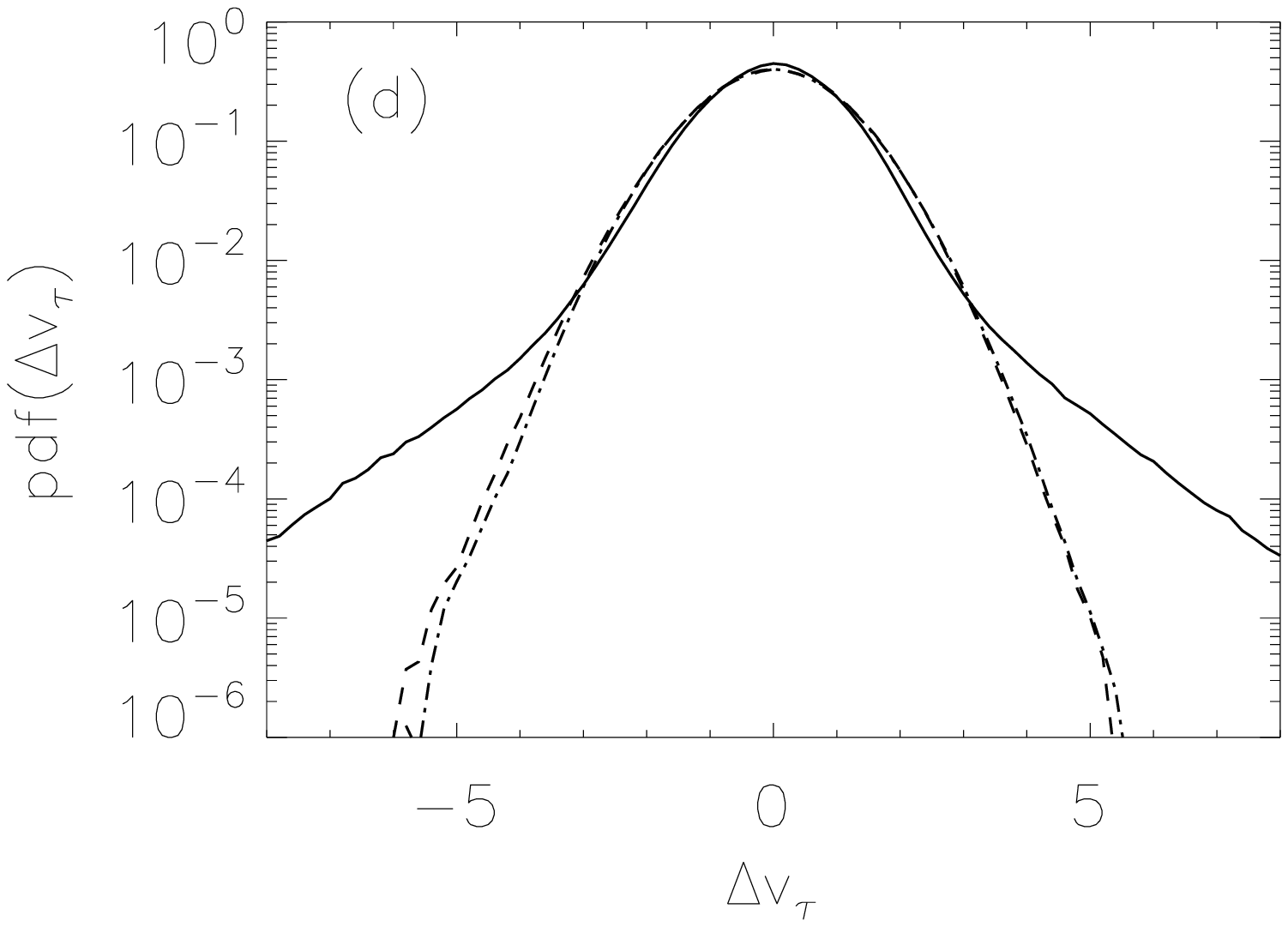}
   \caption{PDFs of line-of-sight velocity increments
   calculated from the \caii\ 854.2 nm line for: (a) the network,
   (b) the fibril, and (c) the internetwork regions. The photospheric PDFs, calculated from
    the Fe\,{\small I} 709.0 nm line over all three regions, are shown in panel (d). 
   The solid, dashed, and dot-dashed lines refer to time separations $\tau=19$~s,
   $\tau=57$~s,and  $\tau=760$~s respectively. Increments are normalized by
   subtracting their mean and dividing by their standard deviation.}
    \label{fig-pdfs}
    \end{figure*}

   \begin{figure}
   \centering
   \includegraphics[width=7.1cm,trim = 15mm 6.1mm 8mm 10.5mm, clip]{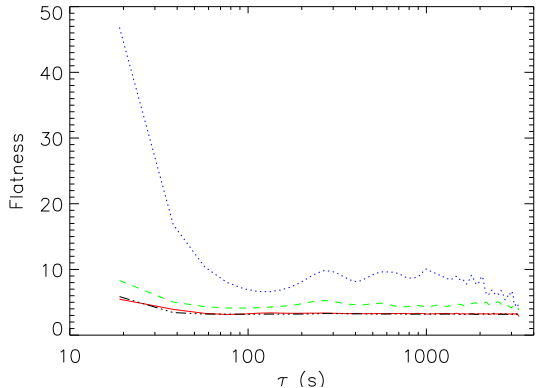}
      \caption{Flatness of line of sight velocity increments as a function of the
increment time separation $\tau$ for the different chromospheric regions as shown
in Fig. \ref{fig:fov} -- the network ({\it dotted line}), fibril ({\it dashed line}), and internetwork
({\it solid line}) regions, as well as for the photospheric line-of-sight velocity increments ({\it dash-dotted line}).}
   \label{fig-flatness}
   \end{figure}

   \begin{figure}
   \centering
   \includegraphics[width=7.1cm,trim = 11mm 5.3mm 9mm 10.2mm, clip]{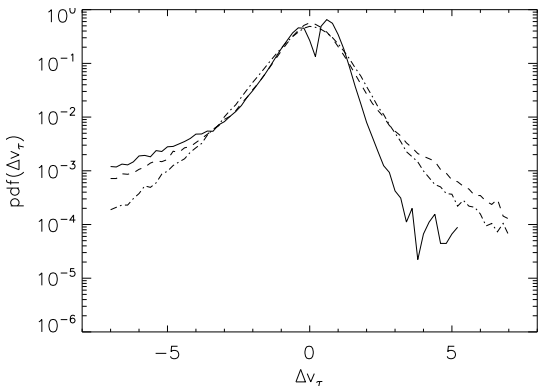}
      \caption{PDFs of chromospheric line-of-sight velocity increments
calculated in the network for
$\tau = 19$~s and
conditioned by the sign of the velocities; the curves refer to 
the cases in which $v(t + \tau )$ and $v(t)$ are both inward ({\it dashed curve}),
both outward ({\it dash-dotted curve}), and of opposite sign ({\it solid curve}). 
Increments are normalized by subtracting their mean and dividing by their
standard deviation.}
         \label{fig-pdfcond}
   \end{figure}

Although the chromospheric power spectra shown in Fig.~\ref{fig:powsp} are not as would be 
expected from simple scaling based on density differences and classical wave propagation, 
they do appear consistent with an energy cascade to smaller scales as found in turbulent flows.
We therefore focus on the statistical properties of large amplitude velocity fluctuations in 
an approach that differs intrinsically from studying solar atmosphere dynamics in terms 
of propagating waves.

Power spectra can completely characterize turbulent fields
only if the parameter fluctuations are distributed according to Gaussian probability density functions
\citep[PDFs; see][]{frisch}. It is thus important in turbulence studies to investigate the 
statistical properties across scale separations $\bvc{\ell}$ to
provide information about the presence of coherent structures, such as vortices or shocks,
at the scale $\bvc{\ell}$.
For  a velocity field $\bvc{v(r)}$ the increments can be defined as
$\Delta \bvc{v_{\ell}} (\bvc{r}) = \bvc{v} (\bvc{r} + \bvc{\ell}) - \bvc{v} (\bvc{r})$ and usually
the longitudinal velocity increments
$\Delta \bvc{v_{\ell_{\parallel}}} (\bvc{r}) = \Delta \bvc{v_{\ell}} (\bvc{r})
\cdot {\bvc{\ell} /  \ell}$ are considered, as in homogeneous and isotropic turbulence
the energy flux can be expressed in terms of the third order moment of longitudinal
velocity increments  \citep[][]{frisch}.
These are also much simpler to measure experimentally if the flow speed in the probe
frame is much larger than the typical velocity fluctuations (e.g., solar wind) which implies that the timescales can be 
transformed into spatial scales according to the Taylor hypothesis \citep[][]{frisch,carbone_rnc}. 
This cannot strictly be assumed in our case, but nevertheless we analyze line-of-sight velocity 
increments as a function of time separation, as this approach is the closest one we can follow with our observational data with respect to the standard analysis of longitudinal velocity increments in turbulence investigations. By denoting by $v(t)$ the line of sight velocity in a given spatial position of the field of view, we consider increments defined by $\Delta v_{\tau}(t) = v(t + \tau ) - v(t)$, where $\tau$ is the time separation.
Figure~\ref{fig-pdfs} shows PDFs of the velocity increments for different time separations, for the 
three chromospheric regions and the photospheric velocities separately. 

The chromospheric PDFs are in all cases non-Gaussian and asymmetric,
with a prevalence of negative fluctuations. The asymmetry increases toward
small scales, with the negative tail becoming more dominant. In comparison, 
the photospheric PDFs are nearly symmetric, but they also
change shape with time separation, being nearly Gaussian at large scales 
but developing tails at small scales. The change in the shape of the PDFs observed in both the
chromosphere and the photosphere indicates a breakdown of self-similarity, which can be
attributed to intermittency phenomena associated with a turbulent cascade. The tails
appear to be more developed in the chromospheric network than in the other regions.

A measure of the intermittency of velocity increments is given by
the flatness (i.e. the ratio of the fourth-order moment to the square of
the second-order moment of the increments). The flatness is 3 for a Gaussian distribution.
The flatness of chromospheric and photospheric
velocity increments as a function of the time lag $\tau$ is shown in Fig.
\ref{fig-flatness}.
The flatness of network velocity increments is always larger than it is in the other
regions, and, most importantly, a significant increase of the flatness at small
scales is found for the network, indicating that intermittency is stronger
in these regions.

In order to have a better statistical characterization of the small-scale velocity
increments in the network, Fig. \ref{fig-pdfcond} shows the PDFs of the velocity increments
conditioned by the sign of the velocities, that is, we calculate the PDFs of
 $\Delta v_{\tau}(t) = v(t + \tau ) - v(t)$  for $\tau = 19$~s separately for the
cases in which $v(t + \tau )$ and $v(t)$ are both positive (inward), both negative (outward), or
each with opposite sign.

For the latter case a local minimum is present for increments
close to 0 that are obviously less probable when the two velocities have opposite
signs. The conditioned PDF analysis indicates that the tail at negative fluctuations
in the network is mainly due to the contribution of structures with inward velocities
or opposite velocities.
 
\section{Conclusions}
\label{sec:conclusions}

We performed Fourier analysis of spatially and temporally resolved measurements 
in the \caii~854.2 nm line. We find that high-frequency velocity fluctuations in the lower 
chromosphere follow a clear power-law behavior, suggesting turbulence 
in the chromospheric plasma. We studied the PDFs of the velocity increments as a function of the time lag
in order to characterize this turbulence for each of the three defined chromospheric 
regions (i.e. network, fibril, and internetwork), which showed stronger intermittency in the network areas.

A high-frequency tail is also seen in the photospheric velocity up to a frequency of almost 15 mHz. This is limited by the higher noise level relative to the observed velocities in the photosphere compared to the chromosphere due to the strong density decrease with height. Our observed photospheric power spectra show a smooth monotonic decrease with frequency, without an increase or structure above the cutoff frequency as predicted, for example, by \citet{1997A&A...324..717T}

Although it is possible that the high-frequency velocities seen in the chromosphere are simply due to the upward propagation of short-period acoustic waves, we find that the chromospheric power spectra and the PDFs are significantly different from the photospheric power spectrum and PDF, indicating that the velocities in the chromosphere at these frequencies are not merely the result of upward-propagating waves. As \citet{2002ESASP.505..293C} point out, only a small percentage of the power present in the photosphere at high-frequencies would be expected to survive up to higher layers because of radiative damping.

We suggest instead that the observed chromospheric turbulence is generated by the acoustic shocks that are present at this height because of the steep vertical density gradients and that the non linear shock processes produce the cascade of energy to higher frequencies. This provides a mechanism for depositing energy into the chromosphere that is more evenly distributed both temporally and spatially than the shock occurrences themselves.  
Since shocks are observed to be abundant (with different characteristics) in both the network and internetwork regions of our data \citep{2008arXiv0807.4966V}, similar turbulent cascades are indeed to be expected in both regions. The resemblance of the network and fibril power spectra may result from the direct connection of the magnetic field from the network into the fibrils.

The presence of non linear turbulent dissipation in the chromosphere is an important consideration in the modeling of this layer of the solar atmosphere. The presence of non linearities and the generation of high-frequency fluctuations there suggest that traditional oscillation analysis, based on the phase and coherence between the photospheric and chromospheric fluctuations, may not properly capture the complexities of the dynamic behavior. Extrapolations of the photospheric energy spectrum to higher layers \citep{2007ApJS..171..520C, 2007ApJ...662..669V} should take into account the modification of the energy spectrum in the chromosphere \citep{1977SoPh...52..163W}; this modification is due to processes occurring close to the acoustic cutoff frequency. 

Finally, the presence of significant mechanical energy in the chromosphere with a broad spectrum of time scales may play an important role in coronal heating \citep{1994ApJ...435..482P}. The observed high-frequency plasma motions are conveniently generated near the surface where the plasma $\beta$ is close to 1. This allows for efficient conversion of these purely acoustic motions into a variety of wave modes, through interaction with the pervasive magnetic field, and provides many additional routes for the continued outward propagation of this acoustic energy. If coronal loops are driven at frequencies comparable to the Alfv\'en crossing time (typically $10\!-\!100$ sec), the amplitude of Alfv\'enic perturbations may increase resonantly and may efficiently dissipate the wave energy in an impulsive fashion \citep{1984ApJ...277..392H, 1997ApJ...490..442M, 2004PhRvL..92s4501N}. 
Acoustic shocks are presumably ubiquitous over the solar surface and may therefore provide an additional source of energy for heating the coronal plasma.

\begin{acknowledgments}
IBIS was constructed by INAF/OAA with contributions from 
the University of Florence, the University of Rome, MIUR, MAE, and is operated with the support of the National Solar Observatory, which is operated by the Association of Universities for Research in Astronomy, Inc., under cooperative agreement with the National Science Foundation.
This work is partially supported by ASI/INAF grant I/015/07/0 ``Studi di Esplorazione del Sistema Solare''. 
We thank Rob Rutten for suggesting many improvements to the presentation of this Letter.
\end{acknowledgments}

\end{document}